 \documentclass[aps,twocolumn,superscriptaddress,showpacs]{revtex4}

\usepackage{epsf}
\usepackage{amsmath}
\usepackage{amssymb}

\def\<{\langle}
\def\>{\rangle}

\newcommand{\tr}{{\rm tr}\,}

\newcommand{\ER}[1]{{E_{R}^{\rm (#1)}}}
\newcommand{\RHO}[1]{{\rho_{\rm #1}}}
\newcommand{\SIGMA}[1]{{\sigma_{\rm #1}}}
\def \rank{{\rm rank\,}}

\begin{document}

\title{Closed formula for the relative entropy of entanglement}

\author{Adam Miranowicz}
\affiliation{Faculty of Physics, Adam Mickiewicz University,
61-614 Pozna\'n, Poland}

\author{Satoshi Ishizaka}
\affiliation{Nano Electronics Research Laboratories, NEC Corporation,
34 Miyukigaoka, Tsukuba 305-8501, Japan}
\affiliation{INQIE, the University of Tokyo,
4-6-1 Komaba, Meguro-ku, Tokyo 153-8505, Japan}

\date{\today}

\begin{abstract}

The long-standing problem of finding a closed formula for the
relative entropy of entanglement (REE) for two qubits is
addressed. A compact-form solution to the inverse problem, which
characterizes an entangled state for a given closest separable
state, is obtained. Analysis of the formula for a large class of
entangled states strongly suggests that a compact analytical
solution of the original problem, which corresponds to finding the
closest separable state for a given entangled state, can be given
only in some special cases. A few applications of the compact-form
formula are given to show additivity of the REE, to relate the REE
with the Rains upper bound for distillable entanglement, and to
show that a Bell state does not have a unique closest separable
state.

\end{abstract}


\maketitle

\pagenumbering{arabic}

%------------------------------------------------------------------
\section{Introduction}

The {\em relative entropy of entanglement} (REE) is as an
entanglement measure quantifying how much a given entangled state
can be distinguished operationally from the set of separable
states or those with positive partial transposition (PPT)
\cite{Vedral97a}:
\begin{equation}
E_R(\rho)={\rm min}_{\sigma' \in {\cal D}} S(\rho ||\sigma' )
=S(\rho ||\sigma), \label{N01}
\end{equation}
where ${\cal D}$ denotes a set of separable states or PPT states,
$S$ is a quasidistance measure usually chosen to be the quantum
relative entropy, $S(\rho ||\sigma' )=\tr\,( \rho \lg \rho
-\rho\lg \sigma' )$, an analog of the classical Kullback-Leibler
divergence. A state $\sigma$ on the boundary of separable states
is called the {\em closest separable state} (CSS) or the closest
PPT state. Various properties of the REE have already been
described (see, e.g.,
\cite{Vedral97a,Vedral98,Horodecki99,Plenio00,Audenaert01,Rains01,Vollbrecht01,Verstraete01,Verstraete01jpa,Verstraete02,Audenaert02,Ishizaka02,Ishizaka03,Ishizaka04,MG04,Wei04}
and for a review see \cite{Horodecki-review}). But there is still
an open fundamental problem \cite{Eisert}: the task of finding a
closed explicit formula for the REE for a given two-qubit state
corresponding to a solution of the convex optimization problem for
the REE, or, briefly, of finding the CSS $\sigma$ for a given
entangled state $\rho$.

Here, we give a compact-form solution of the closely related
problem of finding entangled states $\rho$ and their REE for a
given CSS $\sigma$. Our formula is derived from the results
obtained by one of us in Ref. \cite{Ishizaka03}. We also
demonstrate the intrinsic difficulty in solving the original
problem. In addition, we apply our formula to relate the REE with
the Rains upper bound for the distillable entanglement, and to
show additivity of the REE and nonuniqueness of the CSS for a Bell
state.

%------------------------------------------------------------------
\section{A closed formula for the REE}

Let us consider an entangled two-qubit state $\rho$ with its CSS
$\sigma$, and assume that $\sigma$ is full rank. Note that the
case includes the arbitrary full-rank entangled state $\rho$,
since $\hbox{rank}(\sigma)\ge\hbox{rank}(\rho)$ must hold so that
$S(\rho||\sigma)$ is finite, and hence the CSS $\sigma$ for
full-rank $\rho$ is always full rank. Since $\sigma$ is an edge
state then its partial transposition $\sigma^{\Gamma}$ is rank
deficient, i.e., $\rank(\sigma^{\Gamma})=3$. Let $|\phi\rangle$ be
the kernel (or null space) of $\sigma^\Gamma$, i.e., an eigenstate
of $\sigma^\Gamma$ corresponding to zero eigenvalue,
\begin{eqnarray}
  \sigma^\Gamma |\phi\rangle &=& 0.
\label{N02}
\end{eqnarray}
Moreover, let $|i\rangle$ and $\lambda_i$ be eigenstates and
eigenvalues of $\sigma$, respectively. Using the kernel
$|\phi\rangle$, the formula in Ref. \cite{Ishizaka03} can be
rewritten in the following simple form:
\begin{eqnarray}
  \rho &=& \sigma - x G(\sigma),
  \label{N03}
\end{eqnarray}
where
\begin{eqnarray}
   G(\sigma) &=& \sum_{ij}G_{ij}|i\rangle\langle i|
  (|\phi\rangle\langle \phi|)^\Gamma |j\rangle\langle j|,
 \label{N04} \\
G_{ij}&\equiv& \left\{
\begin{array}{cl}
\lambda_i &\hbox{~~for $\lambda_i\!=\!\lambda_j$,} \cr
\frac{\lambda_i-\lambda_j}{\lg \lambda_i-\lg \lambda_j}
         &\hbox{~~for $\lambda_i\!\ne\!\lambda_j$,}
\end{array}
\right.  \label{N05}
\end{eqnarray}
and $x\ge0$. All $\rho$ obtained from Eq.\ (\ref{N03}) for $x_{\rm
max}\ge x >0$ have $\sigma$ as their CSS, where $x_{\rm max}$ is
the threshold for $\rho\ge0$. This is a unique solution of all
extremal conditions in two qubits as shown later. The same
relation holds for entangled states and the closest PPT state even
in higher-dimensional systems as long as the (PPT) entanglement
witness (EW) $Z$ such that $\tr Z\sigma=0$ is uniquely determined
as $(|\phi\rangle\langle\phi|)^\Gamma$. Moreover, from Eqs.
(\ref{N03})--(\ref{N05}) we have
\begin{eqnarray}
 \langle i |\rho|i\rangle &=& \lambda_i \big[1-x \langle i|
  (|\phi\rangle\langle \phi|)^\Gamma |i\rangle\big].
\label{N06}
\end{eqnarray}
Therefore, the REE can be rewritten as
\begin{eqnarray}
  E_R(\rho) &=& \tr \rho\lg\rho - \tr \rho \lg \sigma
 \nonumber \\
&=& \tr \rho\lg\rho - \sum_i \langle i|
  \rho|i\rangle \lg \lambda_i
\nonumber \\
&=& \tr \rho\lg\rho - \tr \sigma \lg \sigma
\label{N07} \\
&&+ x \sum_i \langle i|
  (|\phi\rangle\langle \phi|)^\Gamma |i\rangle \lambda_i \lg \lambda_i
\nonumber \\
&=& S(\sigma) - S(\rho) + x \tr \big[ (|\phi\rangle\langle
\phi|)^\Gamma \sigma \lg \sigma \big], \nonumber
\end{eqnarray}
where $S(\cdot)$ is the von Neumann entropy. In any case, however,
Eq. (\ref{N03}) should be conversely solved with respect to
$\sigma$ to obtain the true closed formula for the REE as a
solution to Eisert's problem \cite{Eisert}. In the following we
show that the inversion is possible in some special cases but
rather not in general.

Our formula can be derived from the results obtained in Ref.
\cite{Ishizaka03} but the derivation is rather lengthy.
Fortunately, we can also derive it in a compact and elegant way
using a different approach based on the result shown in Ref.
\cite{Rehacek03} that the operator of
\begin{eqnarray}
Z &=& I-\int_0^{\infty} \frac{1}{\sigma+z} \rho \frac{1}{\sigma+z} dz
\label{N08}
\end{eqnarray}
must be an entanglement witness. Let us briefly repeat the proof
in Ref. \cite{Rehacek03} for our later convenience. Since $\sigma$
is the CSS for $\rho$, the inequality
\begin{equation}
S(\rho||(1-\epsilon)\sigma+\epsilon\sigma')-S(\rho||\sigma)\ge0
\label{N09}
\end{equation}
must hold for every separable state $\sigma'$ and
$0\le\epsilon\le1$. Using the expansion of
$$
\lg(X+\epsilon Y)=\lg(X)+\epsilon\int_0^\infty \frac{1}{X+z}Y\frac{1}{X+z}dz
+{\cal O}(\epsilon^2),
$$
we have
\begin{eqnarray}
&&\hspace*{-1cm} S(\rho||(1-\epsilon)\sigma+\epsilon\sigma')-S(\rho||\sigma) \cr
&=& \epsilon\tr \rho\int_0^\infty
\frac{1}{\sigma+z}(\sigma-\sigma')\frac{1}{\sigma+z} dz +{\cal O}(\epsilon^2)\cr
&=& \epsilon \big[\tr \rho-\tr \sigma'\int_0^\infty
\frac{1}{\sigma+z}\rho\frac{1}{\sigma+z}dz\big] +{\cal O}(\epsilon^2)\cr
&=&\epsilon\tr Z\sigma'+{\cal O}(\epsilon^2),
\label{N10}
\end{eqnarray}
and hence $\epsilon\tr Z\sigma'+{\cal O}(\epsilon^2)\ge0$ must
hold for arbitrary small $\epsilon>0$ if $\sigma'$ is a separable
state. This implies that $\tr Z\sigma' \ge0$ must hold for every
separable state $\sigma'$, and therefore $Z$ must be an EW
\cite{Rehacek03}.

Now suppose that $\sigma$ and $\rho$ are two-qubit states. It was
shown in Ref. \cite{Lewenstein00} that an EW in two qubits must be
decomposable, since there are no PPT entangled states in two
qubits. Therefore, $Z$ must be a decomposable EW, and hence
$Z=P+Q^{\Gamma}$, where $P$ and $Q$ are positive operators
\cite{Lewenstein00}. Moreover,
\begin{equation}
\tr Z\sigma = \tr\left[
\sigma-\rho\int_0^{\infty} \frac{1}{\sigma+z} \sigma \frac{1}{\sigma+z} dz
\right]=\tr(\sigma-\rho)=0,
\label{trZs}
\end{equation}
and as a result $\tr Z\sigma=\tr  P\sigma + \tr
Q\sigma^{\Gamma}=0$ must hold. Since $\sigma$ is full rank, and
$\sigma^{\Gamma}$ is positive and rank 3, the solution is uniquely
determined (leaving out the normalization of $x>0$) as $P=0$ and
$Q=x|\phi\rangle\langle \phi|$. As a result,
$x(|\phi\rangle\langle\phi|)^{\Gamma} = Z$ holds for full rank
$\sigma$. The integral in $Z$ can be performed using the
eigenstates $|i\rangle$ for $\sigma$ such that
\begin{eqnarray}
x\langle i|(|\phi\rangle\langle\phi|)^\Gamma|j\rangle
&=&\delta_{ij}-\int_0^\infty\frac{1}{\lambda_i+z}\langle i|\rho|j\rangle
\frac{1}{\lambda_j+z}dz \cr
&=&\delta_{ij}-\langle i|\rho|j\rangle G_{ij}^{-1},
\end{eqnarray}
and finally we have Eq. (\ref{N03}). In this way, the satisfaction
of all extremal conditions in the optimization problem for the REE
is automatically ensured by the condition that $Z$ is an EW such
that $\tr Z\sigma =0$. Note that $|\phi\rangle$ is always
entangled for full rank $\sigma$ \cite{Verstraete01b} (if
$|\phi\rangle$ is not entangled, the full rank $\sigma$ is not a
CSS for any entangled state).

%------------------------------------------------------------------
\section{The Rains bound and the REE}

An upper bound for distillable entanglement introduced by Rains \cite{Rains01}
is defined as
\begin{equation}
R(\rho)=\min_{\tau'\ge0}\big[ S(\rho ||\tau' )+\lg\tr |\tau'^\Gamma|\big],
\label{N11}
\end{equation}
where minimization is taken over all states including entangled states,
and hence $R(\rho)\le E_R(\rho)$ follows from the definition.
Here, let us apply the technique as used in the previous section to
the optimization problem for the Rains bound.
It was shown in Ref. \cite{Audenaert02} that the problem is
reduced to
\begin{equation}
R(\rho)=\min_{\tau'\ge0} S(\rho||\tau'),
\label{N12}
\end{equation}
where the minimization is taken over {\em unnormalized} states
subject to $\tr |\tau'^\Gamma|\le1$. This is a convex optimization
problem because $\tr |\tau_0^\Gamma|\le 1$ for
$\tau_0=p\tau_1+(1-p)\tau_2$ and
$\tr |\tau_1^\Gamma|,\tr |\tau_2^\Gamma|\le1$ \cite{Audenaert02}.
Suppose that $\rho$ is full rank and $\tau$ is an optimal unnormalized state.
Hence, $\tau$ is full rank, $\tr |\tau^\Gamma|=1$, and
\begin{equation}
S(\rho||(1-\epsilon)\tau+\epsilon\tau')-S(\rho||\tau)\ge0
\label{N13}
\end{equation}
for every $\tau'$ such that $\tr |\tau'^\Gamma|\le1$
and $0\le\epsilon\le1$.
Let $\tau'$ be a normalized separable state, i.e., $\tau'^\Gamma\ge0$ and
$\tr |\tau'^\Gamma|=\tr \tau'=1$.
Using the expansion of the logarithmic function
for $\epsilon\rightarrow+0$ as in the previous section, we
have
\begin{eqnarray}
&&\hspace*{-1cm} S(\rho||(1-\epsilon)\tau+\epsilon\tau')-S(\rho||\tau) \cr
&\sim&\epsilon\tr \big[
I-\int_0^\infty
\frac{1}{\tau+z}\rho\frac{1}{\tau+z}dz\big]\tau' \cr
&\equiv&\epsilon\tr Z_R\tau' \ge 0,
\label{N14}
\end{eqnarray}
and hence $Z_R$ must be again an EW. Note however that $\tr
Z_R\tau=\tr \tau-\tr \rho\le0$ in this case, contrary to Eq.\
(\ref{trZs}), because $\tau$ is unnormalized so that $\tr
|\tau^\Gamma|=1$.

Let us then consider the case where $\rho$ is a two-qubit state,
and suppose that the optimal two-qubit state $\tau$ is entangled.
Since the partial transposition of a two-qubit state has only one
negative eigenvalue \cite{Verstraete01jpa}, $\tau^\Gamma$ is
expressed such that
\begin{equation}
\tau^\Gamma=(1-\mu)\Pi-\mu|\psi\rangle\langle\psi|,
\label{N15}
\end{equation}
where $\Pi\ge0$, $\Pi|\psi\rangle=0$, and $\mu>0$. Moreover, $\tr
\Pi=1$ so that $\tr |\tau^\Gamma|=1$. For a small deviation of
$\mu\rightarrow(1+\delta)\mu$, we have
\begin{eqnarray}
S(\rho||\tau)&\rightarrow&
S(\rho||\tau)-
\delta\mu\big[\tr Z_R\big(\Pi^\Gamma+(|\psi\rangle\langle\psi|)^\Gamma\big)-2\big] \cr
&=&S(\rho||\tau)-
\delta\big[\tr Z_R(\Pi^\Gamma-\tau)-2\mu\big] \cr
&=&S(\rho||\tau)-
\delta \tr Z_R\Pi^\Gamma ,
\label{N16}
\end{eqnarray}
where $\tr Z_R\tau=\tr \tau-1=-2\mu$ was used. Since $\tau$ must
satisfy the extremal condition with respect to $\mu$, $\tr
Z_R\Pi^\Gamma =0$ must hold. Moreover, since $\tau$ is a two-qubit
entangled state, $\Pi$ is rank-3 and $\Pi^\Gamma$ is positive definite
\cite{Ishizaka04}, and as a result the EW $Z_R$ is uniquely
determined (including the normalization in this case) as
$Z_R=2(|\psi\rangle\langle\psi|)^\Gamma$, where $\tr
Z_R\tau=-2\mu$ and $\mu\ne0$ were taken into account. This implies
that
\begin{equation}
I-2(|\psi\rangle\langle\psi|)^\Gamma=\int_0^\infty \frac{1}{\tau+z}\rho\frac{1}{\tau+z}dz,
\label{N17}
\end{equation}
but this cannot be satisfied because the right-hand side is
positive definite for full-rank $\rho$ and $\tau$, while the
left-hand side is not for any $|\psi\rangle$. Therefore, the
optimal state $\tau$ must not be entangled, and the optimization
in $R(\rho)$ is achieved by a separable state. The same discussion
also holds for low-rank $\rho$, because $R(\rho)$ is a continuous
function \cite{Audenaert02}. It is then concluded that
$R(\rho)=E_R(\rho)$ for every two-qubit state, and our
compact-form formula also holds for the Rains bound.

Note that the Rains bound is strictly smaller than the REE for the
Werner state in higher-dimensional systems
\cite{Rains01,Audenaert01}, but such a disagreement does not occur
in two qubits as shown above.

%------------------------------------------------------------------
\section{Additivity of the REE}

An asymptotic REE defined as
\begin{equation}
E^\infty_R(\rho)=\lim_{n\rightarrow\infty}\frac{1}{n}E_R(\rho^{\otimes
n}) \label{N18}
\end{equation}
satisfies $E^\infty_R(\rho)\le E_R(\rho)$ from the definition.
The equality holds if $E_R(\rho)$ is weakly additive, but this is not the case
in general \cite{Audenaert01}. Here, let us briefly investigate the additivity
using our compact-form formula.

In Ref. \cite{Rains}, it was shown that $E_R(\rho)$ for $\rho$
such that $[\rho,\sigma] = 0$ is weakly additive if
$(\rho\sigma^{-1})^{\Gamma}\ge - \openone$. From Eq. (\ref{N03}),
\begin{eqnarray}
&&\hspace*{-1.5cm}\langle i|(\lg \lambda_i \rho-\rho\lg \lambda_j)|j\rangle \cr
&=&-x\langle i|\left[ \lambda_i (|\phi\rangle\langle\phi|)^\Gamma
- (|\phi\rangle\langle\phi|)^\Gamma\lambda_j \right]|j\rangle
\end{eqnarray}
must hold for all $i$ and $j$, and hence we have $[\lg
\sigma,\rho]=-x[\sigma,(|\phi\rangle\langle\phi|)^\Gamma]$. This
implies that $[\rho,\sigma] = 0$ if and only if
$[\sigma,(|\phi\rangle\langle \phi|)^\Gamma] = 0$. Therefore,
$(|\phi\rangle\langle \phi|)^\Gamma$ must be diagonalized in terms
of the eigenstates of $\sigma$ so that $[\rho,\sigma] = 0$, and
the compact-form formula in this case is much simplified as
\begin{eqnarray}
  \rho &=&\sigma- x(|\phi\rangle\langle \phi|)^\Gamma\sigma.
\label{N19}
\end{eqnarray}
Let $p_0 \ge 1/2$ be the maximal Schmidt coefficient of
$|\phi\rangle$. Since the largest eigenvalue of
$(|\phi\rangle\langle \phi|)^\Gamma$ is $p_0$, the range of $x$
must satisfy $0 \le x \le 1/p_0\le 2$, so that $\rho \ge 0$. As a
result, $(\rho\sigma^{-1})^{\Gamma}= I-x |\phi\rangle\langle\phi|
\ge -\openone$ always holds. Therefore, it is concluded that
$E_R(\rho)$ is weakly additive and $E^\infty_R(\rho)=E_R(\rho)$
for every two-qubit state such that $[\rho,\sigma] = 0$.

Moreover,  it was shown in Ref. \cite{Rains} that $E_R(\rho)$ for
$\rho$ such that $[\rho,\sigma] = 0$ is strongly additive, namely
$E_R(\rho\otimes \tau)=E_R(\rho)+E_R(\tau)$ for an arbitrary
$\tau$, if $(\rho\sigma^{-1})^{\Gamma}\ge 0$. In the same way as
above, it is found that the condition is satisfied if $x\le 1$,
and therefore $E_R(\rho)$ is strongly additive for every two-qubit
state such that $[\rho,\sigma] = 0$ and $x\le 1$.

%------------------------------------------------------------------
\section{Formula applications for full-rank CSS}

All the examples of arbitrary rank states $\rho$ with their CSS
$\sigma$ found by us in the literature
\cite{Vedral97a,Vedral98,Verstraete01,Verstraete01jpa,Verstraete02,Audenaert02,MG04,Wei04},
can easily be explained using our formula. The procedure can be
summarized as follows: choose a full-rank matrix $\sigma$,
calculate its partial transposition to get $\sigma^{\Gamma}$, find
a condition on its elements for which $\sigma^{\Gamma}$ is rank
deficient (and so becomes a CSS), calculate
$\rho=\sigma-xG(\sigma)$, if required take a limit of some
elements to diminish the rank of $\sigma$ and $\rho$, and finally
find an inverse relation to express the elements of $\sigma$ in
terms of those of $\rho$.

For example, let us analyze a full-rank state $\sigma=\sum_{i=1}^4
R_i |\beta_i\>\<\beta_i|$ diagonal in the Bell basis
$\{|\beta_i\>\}$. The eigenvalues of $\sigma^{\Gamma}$ are
$\Lambda_i=\frac12-R_i$. Thus, e.g., by setting $\Lambda_1=0$,
$\sigma$ becomes a CSS. By noting that the kernel $|\phi\>$ is a
Bell state and applying it to Eq. (\ref{N03}), one gets a
Bell-diagonal entangled state $\RHO{BD}=\sum_{i} r_i
|\beta_i\>\<\beta_i|$, where $r_1=(2+x)/4$ and otherwise
$r_i=R_i(1-x/2)$. By inverting the latter equation, one gets the
well-known formula \cite{Vedral97a}
\begin{eqnarray}
  \SIGMA{BD} &=& \frac12 |\beta_1\>\<\beta_1| +
  \frac{1}{2(1-r_1)}\sum_{i=2}^4 r_i |\beta_i\>\<\beta_i|.
\label{N20}
\end{eqnarray}
This is the CSS of an arbitrary Bell-diagonal state
$\RHO{BD}=\sum_{i=1}^4 r_i |\beta_i\>\<\beta_i|$ assuming $r_1\ge
1/2$.

As another example which, to our knowledge, has not been discussed
in the literature, let us analyze a two-qubit state of the
following form:
\begin{eqnarray}
\SIGMA{Z}=
\begin{pmatrix}
R_1 & 0   & 0   & 0 \\
  0 & R_2 & Y   & 0 \\
  0 & Y   & R_3 & 0 \\
  0 & 0   & 0   & R_4
\end{pmatrix}.
\label{N21}
\end{eqnarray}
This state is the CSS if its partial transposition $\sigma_{\rm
Z}^{\Gamma}$ is rank 3, which implies that
$Y=\sqrt{R_1R_4}e^{i\varphi}$ (in the following we set
$\varphi=0$), while the requirement of positivity of the density
operator $\SIGMA{Z}$ implies that $R_2R_3\ge Y^2$. Thus, $\RHO{Z}$
satisfies the condition $R_2R_3\ge R_1R_4$, and (\ref{N21}) can
compactly be given by
\begin{eqnarray}
  \SIGMA{Z} &=& {\cal N} (|\psi\>\<\psi|)^{\Gamma}+R_2 |01\>\<01|
  +R_3 |10\>\<10|,
\label{N22}
\end{eqnarray}
where $|\psi\>={\cal N}^{-1/2}(\sqrt{R_1}|00\>+\sqrt{R_4}|11\>)$
and ${\cal N}=R_1+R_4$. The eigenvalues of $\SIGMA{Z}$ are
$(\lambda_i)_i=(R_1,R_4,\lambda_+,\lambda_-)$, where
$\lambda_{\pm}=\frac12(R_2+R_3\pm z)$ with the auxiliary function
$z=\sqrt{(R_2-R_3)^2+4Y^2}$. The corresponding eigenvectors are
$|\lambda_1\>=|00\>$, $|\lambda_2\>=|11\>$, and
$|\lambda_{\pm}\>={\cal N}_{\pm}[(\lambda_{\pm}-R_3)|01\>+Y|10\>]$
with normalizations ${\cal
N}_{\pm}=[(\lambda_{\pm}-R_3)^2+Y^2]^{-1/2}$. One finds the kernel
$|\phi\>$ of $\sigma_{\rm Z}^\Gamma$ and then
\begin{eqnarray}
(|\phi\>\<\phi|)^{\Gamma} \!=(R_1+R_4)^{-1}(R_4|00\>\<00|-
Y|01\>\<10|
\nonumber \\
-Y|10\>\<01|+R_1|11\>\<11|).\label{N22a}
\end{eqnarray}
Thus, according to (\ref{N04}), we find that
\begin{eqnarray}
G(\SIGMA{Z})=
\begin{pmatrix}
\bar R_1 & 0   & 0   & 0 \\
  0 & \bar R_2 & \bar Y   & 0 \\
  0 & \bar Y   & \bar R_3 & 0 \\
  0 & 0   & 0   & \bar R_4
\end{pmatrix},
\label{N23}
\end{eqnarray}
where
\begin{eqnarray}
  \bar R_1&=&\bar R_4=\frac{Y^2}{R_1+R_4},
\nonumber \\
  \bar R_2&=& -2 Y^2 d [(R_2-R_3)(z+R_2 L)+2 Y^2 L],
\nonumber \\
   \bar R_3&=& -2\bar R_1-\bar R_2,
\nonumber \\
  \bar Y &=& Y d [2 Y^2 (R_2+R_3) L- (R_2-R_3)^2z]
 \label{N24}
\end{eqnarray}
are given in terms of
\begin{eqnarray}
  z&=&\sqrt{(R_2-R_3)^2+4R_1R_4},
\nonumber \\
  L&=&\ln(R_2+R_3-z)-\ln(R_2+R_3+z),
   \label{N25}
\end{eqnarray}
and $1/d=(R_1+R_4)z^2 L$.

Thus, according to (\ref{N03}), the entangled states
\begin{eqnarray}
  \RHO{Z} &=& \SIGMA{Z} - x G(\SIGMA{Z})
\label{N26}
\end{eqnarray}
have the same CSS $\SIGMA{Z}$. This is an important example in our
analysis as all low-rank states discussed in Sec. VI are special
cases of (\ref{N26}). Note that it is required to assume $x\le
x'_{\max}=(R_1+R_4)/R_1$ to ensure that $(R_4-x\bar R_4)\ge 0$.
Assuming for simplicity that $R_1\ge R_4$, the analogous
conditions for $(R_i-x\bar R_i)\ge 0$ with $i=1,2,3$ are also
satisfied for $x\le x'_{\max}$. On the other hand, the condition
$(R_2-x\bar R_2)(R_3-x\bar R_3)\ge (Y-x\bar Y)^2$, which is also
implied by the positivity of $\RHO{Z}$, restricts $x$ to be
smaller than $x''_{\max}=f-\sqrt{f^2-4\Delta\bar \Delta}/(2\bar
\Delta)$, where $\Delta=R_2R_3-Y^2$, $\bar \Delta=\bar R_2\bar
R_3- \bar Y^2$, and $f=R_2\bar R_3+\bar R_2R_3-2Y\bar Y$. Thus,
(\ref{N26}) is defined for $0< x\le x_{\max}\equiv
\min\{x'_{\max}, x''_{\max}\}$. The problem of expressing
$\SIGMA{Z}$ in terms of $\RHO{Z}$ will be addressed in the
following section.

%------------------------------------------------------------------
\section{Formula applications for lower-rank CSS}

Our compact-form formula can also be applied for lower-rank CSSs
$\sigma$ in two approaches: directly for some special states and
indirectly for arbitrary states.

To justify a direct application, the following conditions should
be satisfied: (i) There must exist a full-rank edge state
$\sigma'$ in the vicinity of $\sigma$. If it is certain that
$\sigma$ is a CSS for some $\rho$, this condition is trivial.
However, if we do not know whether or not $\sigma$ can be a CSS
for some $\rho$, this is not trivial. (ii) Let $|\phi'\>$ be the
kernel of $\sigma'^{\Gamma}$, i.e., $\sigma'^{\Gamma}|\phi'\>=0$.
Then, $|\phi'\>$ must be entangled. However, when $|\phi'\>$ is
not entangled, we merely cannot find any entangled $\rho'\ge 0$
for $\sigma'$ by the compact-form formula, and hence this
condition is not so important. (iii) There must exist a sequence
such that $\sigma'\rightarrow \sigma$ and $|\phi'\> \rightarrow
|\phi\>$ simultaneously. This condition seems to severely
constrain the choice of $|\phi\>$ in the case of
rank($\sigma^{\Gamma}$) = 2.

Thus, our formula can be applied directly to the rank-2 Horodecki
state defined for $p\in\langle 0,1 \rangle$ by
\cite{Horodecki-book}
\begin{equation}
\RHO{H}=p|\psi^{(\pm)}\rangle \langle
\psi^{(\pm)}|+(1-p)|00\rangle \langle 00|, \label{N27}
\end{equation}
which is a mixture of a Bell state $|\psi^{(\pm)}\rangle=
(|01\rangle\pm |10\rangle)/\sqrt{2}$ and a separable state
orthogonal to it. It is worth noting that the Horodecki state is
extremal in the sense that it minimizes the REE for a given
concurrence \cite{Verstraete01jpa}, negativity (i.e., a measure of
the PPT entanglement cost) for a given concurrence
\cite{Verstraete01jpa}, fidelity (i.e., maximal singlet fraction)
for a given concurrence ($\ge 1/3$), and negativity [$\ge
(\sqrt{5}-2)/3$] \cite{Verstraete02b}. The state also satisfies
some extremal conditions for the REE for a given negativity
\cite{MG04}. The CSS for $\RHO{H}$ derived from
(\ref{N03})--(\ref{N05}) reads as
\begin{equation}
  \SIGMA{H} = q'^2 |00\rangle
\langle 00|+2p'q'
|\psi^{(\pm)}\rangle\langle\psi^{(\pm)}|+p'^2|11\rangle \langle
11|, \label{N28}
\end{equation}
where $p'=p/2$ and $q'=1-p'$, in agreement with the known result
derived in another way \cite{Vedral98}. Note that, although
$\rank(\sigma_{H}^{\Gamma})=3$, $\SIGMA{H}$ is not full rank.
Fortunately, the above-mentioned conditions necessary for the
direct application of the compact-form formula are satisfied.

In the second more general approach, one can apply our formula for
arbitrary lower-rank states in a limiting sequence from a
full-rank state because the REE is a continuous function. For
example of such application of our formula for lower-rank states
we will analyze pure states and the rank-2 Vedral-Plenio state
defined by \cite{Vedral98}:
\begin{equation}
\RHO{VP}=p|\psi^{(+)}\rangle \langle \psi ^{(+)}|+(1-p)|01\rangle
\langle 01| \label{N29}
\end{equation}
for $0\le p\le 1$ with the corresponding CSS
\begin{equation}
\SIGMA{VP}=\left(1-\frac{p}2\right)|01\rangle\langle 01|
+\frac{p}2|10\rangle\langle 10|, \label{N30}
\end{equation}
for which $\rank(\SIGMA{VP})=\rank(\sigma_{\rm VP}^{\Gamma})=2$.
By contrast with the Horodecki state, (\ref{N29}) is a mixture of
a Bell state and a separable state {\em not} orthogonal to it.

To derive CSSs $\SIGMA{P}$ and $\SIGMA{VP}$, and thus to show the
usefulness of our formula also for lower-rank states, let us apply
state $\SIGMA{Z}$, given by (\ref{N21}), in the limiting cases.
Namely, by assuming in (\ref{N26}) that $x=x'_{\max}\le
x''_{\max}$ and $R_1\ge R_4$, one gets the following extremal
state:
\begin{eqnarray}
\rho'_{\rm Z}\equiv \RHO{Z}(x=x'_{\max})=
\begin{pmatrix}
r_1 & 0   & 0   & 0 \\
  0 & r_2 & y   & 0 \\
  0 & y   & r_3 & 0 \\
  0 & 0   & 0   & 0
\end{pmatrix},
\label{N31}
\end{eqnarray}
where
\begin{eqnarray}
  r_1 &=& R_1-R_4,
\nonumber \\
  r_2 &=&R_2+\frac{2R_4}{z^2}(R_2^2-R_2R_3+2Y^2)+\frac{2R_4}{Lz}(R_2-R_3),
\nonumber \\
  r_3 &=&1-r_1-r_2,
 \label{N32} \\
  y&=&\frac{1}{2Y}[2(R_1+R_2)R_4-(r_2-R_2)(R_2-R_3)].
\nonumber
\end{eqnarray}
In the special case of $\sigma'_{\rm Z}$ for $R_1=R_4\rightarrow
0$, one gets
\begin{eqnarray}
 \sigma''_{\rm Z} &=& (1-R_3)|01\>\<01|+R_3 |10\>\<10|,
\label{N33}
\end{eqnarray}
then $\rho''_{\rm Z}=\sigma''_{\rm Z}-x\bar
Y(|01\>\<10|+|10\>\<01|)$ with $\bar Y= -(1-2R_3)/[4\,{\rm
atanh}(1-2R_3)]$, where atanh is the inverse hyperbolic tangent.
This state for $x= R_3/{\bar Y}$ assuming $R_3\le 1/2$ corresponds
to a generalized Vedral-Plenio state:
\begin{eqnarray}
  \RHO{GVP} &=& p|\psi_P\rangle\langle \psi_P|+ (1-p) |01\rangle\langle 01|,
  \label{N34}
\end{eqnarray}
where in comparison with the standard Vedral-Plenio state
$\RHO{VP}$, given by (\ref{N29}), a Bell state is replaced by a
pure state
\begin{eqnarray}
 |\psi_P\> =
\sqrt{P}|01\>+e^{i\varphi}\sqrt{1-P}|10\>, \label{N35}
\end{eqnarray}
for any $0\le P\le 1$. For simplicity we set $\varphi=0$. Thus,
the CSS $\SIGMA{GVP}$ is just given by Eq. (\ref{N33}) for
$R_3=p(1-P)$. By assuming $R_3=p/2$ and $x= R_3/{\bar R}$, one
gets $\RHO{VP}$ and its CSS $\SIGMA{VP}$, given by (\ref{N30}), in
agreement with the solution obtained in Ref. \cite{Vedral98}.
Moreover, any two-qubit pure state $|\psi\> = c_0|00\>+ c_1|01\>+
c_2|10\>+ c_3|11\>$ can be transformed by local rotations into the
state (\ref{N35}). Thus, the CSS $\sigma''_{\rm Z}$, given by
(\ref{N33}), also describes the CSS $\SIGMA{P}$ for an arbitrary
pure state $|\psi\>$ as expected \cite{Vedral97a}.

A mixed state introduced by Gisin \cite{Gisin96},
\begin{eqnarray}
  \RHO{G} &=& q |00\rangle\langle 00|+p|\psi_P\rangle\langle \psi_P|
  +q|11\rangle\langle 11|,
\label{N36}
\end{eqnarray}
where $|\psi_P\>$ is given by (\ref{N35}), $0 \le p\le 1$, and
$q=(1-p)/2$. By contrast to the generalized Vedral-Plenio state
and the generalized Horodecki state, (\ref{N36}) is a mixture of
an entangled pure state and two separable states orthogonal to it.
The Gisin state is also a special case of $\RHO{Z}$ assuming
$R_1-x\bar R_1=R_4-x\bar R_4=q$. Thus, its CSS is equal to
$\SIGMA{Z}$, given by (\ref{N21}) for $R_1=R_4$. It is worth
noting that (\ref{N36}) is one of the simplest examples of an
entangled state, which does not violate any Bell-type inequality
(for some range of $p$ for a given $P$) \cite{Peres96}.

Assuming $y^2=r_2r_3$, the state $\rho'_{\rm Z}$ reduces to a
rank-2 state, which we refer to as the generalized Horodecki state
defined as
\begin{eqnarray}
  \RHO{GH} &=& p|\psi_P\rangle\langle \psi_P|+ (1-p) |00\>\<00|,
\label{N37}
\end{eqnarray}
where $p=1-r_1$ and $|\psi_P\rangle$ is given by (\ref{N35}) for
$P=r_2/(1-r_1)$. Note that, for $P=1/2$, which corresponds to
$r_2=r_3$, $\RHO{GH}$ reduces to the standard Horodecki state
$\RHO{H}$ given in terms of a Bell state
$|\psi_P\>=|\psi^{(+)}\>$. On the other hand, the state $\RHO{GH}$
for $r_1=(1+C)/2$ and $r_2=d_+$ reduces to the
Verstraete-Verschelde state defined for $C\le 1/3$ by
\cite{Verstraete02b}
\begin{eqnarray}
\RHO{V}=
  \left(
\begin{array}{llll}
 \frac{1+C}{2} & 0 & 0 & 0 \\
 0 & d_+ & \frac{C}{2} & 0
   \\
 0 & \frac{C}{2} & d_- & 0
   \\
 0 & 0 & 0 & 0
\end{array}
\right), \label{N38}
\end{eqnarray}
where $d_{\pm}=\frac{1}{4} \left(1-C\pm\sqrt{1-2 C-3 C^2}\right)$
and $C\equiv C(\RHO{V})$ is the concurrence, which can be
expressed in terms of negativity $N=N(\RHO{V})$, as
$C=\frac12[N+\sqrt{N(4+5N)}]$ holds. It is worth noting that state
(\ref{N38}) minimizes fidelity for a given concurrence and
negativity \cite{Verstraete02b}.

It is seen by analyzing $\rho'_{\rm Z}$ as a function of elements
of $\sigma'_{\rm Z}$ that it seems impossible to invert the
general Eqs. (\ref{N31}) and (\ref{N32}). However, the equations
can be inverted in some special cases. For example, by assuming
$R_2=R_3$ for the state $\rho'_{\rm Z}$, which implies
$r_2=r_3=(1-r_1)/2$, one finds that the general relation for $r_2$
in Eqs. (\ref{N32}) reduces to $r_2=R_2+R_4=1-R_1-R_2$. Under this
assumption, the set of equations (\ref{N32}) can be solved for
$\{R_i\}$ in terms of $\{r_i\}$ and $y\le r_2$ as follows:
\begin{eqnarray}
  R_4&=&\frac{4r_1y^2}{(1+r_1)^2-4y^2},
\nonumber \\
  R_2&=&R_3=r_2-R_4,
\nonumber \\
   R_1&=&r_1+R_4.
\label{N39}
\end{eqnarray}
This state, in the special case of $r_2=y$, reduces to the
standard Horodecki state $\RHO{H}$, given by (\ref{N27}). However,
it does not reduce to the generalized Horodecki state $\RHO{GH}$
if $P\neq \frac12$, for which we can give only a formal solution
for the REE:
\begin{equation}
  \ER{GH} = -H_2(r_1)-r_1 \log R_1-f_-^2 \log \lambda_- -f_+^2 \log \lambda_+,
\label{N40}
\end{equation}
where $H_2(\cdot)$ is the binary entropy and $f_{\pm}={\cal
N}_{\pm}[(\lambda_{\pm}-R_3)\sqrt{r_2}+Y\sqrt{r_3}]$, while
$\lambda_{\pm}$ and ${\cal N}_{\pm}$ are defined below Eq.
(\ref{N22}). A compact-form explicit formula for the REE for the
states $\sigma'_{\rm Z}$ with elements given by (\ref{N39}) as
well as for the Horodecki state or the Vedral-Plenio state is thus
obtained. However, it seems impossible to invert (\ref{N32}) in
order to express all $\{R_i\}$ in terms of $\{r_i\}$ and $y$, even
for the generalized Horodecki state with $P,p\neq 0,\frac12,1$.

%------------------------------------------------------------------
\section{Nonuniqueness of the CSS for Bell states}

Analysis of our formula and the above examples enables us to find
that for a given entangled state there is not always a unique CSS
due to a limiting procedure. For this purpose let us derive CSSs
for a Bell state $|\psi^{(+)}\>$ from $\SIGMA{Z}$, given by
(\ref{N21}). First, assume that $R_1=R_4=\epsilon$ and
$R_2=R_3=1/2-\epsilon$ for small $\epsilon \ge 0$. By taking the
limit of $\epsilon\rightarrow 0$, one gets the following CSS:
\begin{eqnarray}
  \sigma'_{\rm Bell} &=& \lim_{\epsilon\rightarrow 0} \SIGMA{Z} =
  \frac12 (|01\>\<01|+|10\>\<10|),
\label{N41}
\end{eqnarray}
for which $G(\sigma'_{\rm Bell})=-\frac14(|01\>\<10|+|10\>\<01|)$.
By noting that $x_{\rm max}=2$, one finds that $\sigma'_{\rm
Bell}-x_{\rm max}G(\sigma'_{\rm Bell})$ corresponds to the Bell
state $|\psi^{(+)}\>$. The same CSS is obtained from special cases
of the CSS for pure state and the Vedral-Plenio state, given by
(\ref{N30}) for $p=1$. On the other hand, let us analyze
$\SIGMA{Z}$ assuming $R_{i}=1/4 - \epsilon$ for $i=1,...,4$. Then,
in the limit of $\epsilon\rightarrow 0$, $\SIGMA{Z}$ reduces to
the CSS:
\begin{equation}
  \sigma''_{\rm Bell} = \lim_{\epsilon\rightarrow 0} \SIGMA{Z} =
  \frac14 \left(|00\>\<00|+2|\psi^{(+)}\>\<\psi^{(+)}|
  +|11\>\<11|\right), \label{N42}
\end{equation}
for which state $\sigma''_{\rm Bell}-x_{\rm max}G(\sigma''_{\rm
Bell})$ also corresponds to the Bell state $|\psi^{(+)}\>$. Note
that (\ref{N42}) is a special case of the CSS for the Horodecki
state, given by (\ref{N28}) for $p=1$, and for the Gisin state,
given by (\ref{N36}) for $p=P=\frac12$.

In a more general approach, let us analyze a Bell-diagonal state
of the form
\begin{equation}
  \RHO{BD} =
  (1-k\epsilon)|\beta_1\>\<\beta_1|+\epsilon \sum_{i=2}^4
k_i|\beta_i\>\<\beta_i|, \label{N43}
\end{equation}
where $0\le k_i < \infty$ and $k\equiv k_2+k_3+k_4$ such that
$k\epsilon\le 1$. The state in the limit of $\epsilon\rightarrow
0$ reduces to the Bell state $|\beta_1\>$, for which the CSS
depends on $\{k_i\}$ as follows:
\begin{eqnarray}
 \SIGMA{Bell} &=& \frac 12 |\beta_1\>\<\beta_1|+ \frac{1}{2k}\sum_{i=2}^4
k_i|\beta_i\>\<\beta_i| \label{N44}
\end{eqnarray}
according to (\ref{N20}). Thus, an arbitrary Bell-diagonal state
with one of its eigenvalues equal to 1/2 is the CSS for a Bell
state. In special cases, $\SIGMA{Bell}$ goes to (\ref{N41}) for
$k_2=1,k_3=k_4=0$ and to (\ref{N42}) for $k_2=0,k_3=k_4=1$
assuming $|\beta_{1,2}\>=|\psi^{(\pm)}\>$ and
$|\beta_{3,4}\>=|\phi^{(\pm)}\>=\frac{1}{\sqrt{2}}(|00\>\pm
|11\>)$. Other CSSs of Bell states, which are not diagonal in the
Bell basis, can be obtained by rotating $\SIGMA{Bell}$. For
example, the CSS
\begin{equation}
  \sigma'''_{\rm Bell} =
  \frac14\left(2|\psi^{(+)}\>\<\psi^{(+)}|
  +|\psi\>\<\psi|
  +|\phi^{(-)}\>\<\phi^{(-)}|\right),
\label{N45}
\end{equation}
where $|\psi\>=\frac{1}{\sqrt{2}}(|\psi^{(-)}\>+|\phi^{(+)}\>)$,
is obtained by rotating $\sigma''_{\rm Bell}$ given by
(\ref{N42}).

It is worth noting that the generated Bell-diagonal state
$\rho=\SIGMA{Bell}-xG(\SIGMA{Bell})$ is independent of the
parameters $\{k_i\}$ in $\SIGMA{Bell}$ only for $x=x_{\rm max}$,
but it depends on the choice of $\{k_i\}$ for $x<x_{\rm max}$,
although the largest eigenvalue of $\rho$, $\lambda_1=\max({\rm
eig}\rho)$, is $\{k_i\}$ independent. Thus, the REE is also
independent of $\{k_i\}$, as $E_{\rm R}(\rho)=1-H_2(\lambda_1)$
for $\lambda_1\ge \frac12$ and $E_{\rm R}(\rho)=0$ otherwise.
Similarly, other entanglement measures, including the negativity
and concurrence are $\{k_i\}$ independent, as
$N(\rho)=C(\rho)=2\lambda_1-1$. By contrast, violation of
Bell-inequality by $\rho$ depends on all eigenvalues
$\{\lambda_i\}$, so it depends on the choice of $\{k_i\}$ for
$\SIGMA{Bell}$. This can be seen explicitly, by analyzing the
Horodecki parameter $M$ \cite{Horodecki95,Miran} describing a
degree of the violation of the Bell inequality due to Clauser,
Horne, Shimony, and Holt.

%------------------------------------------------------------------
\section{Conclusion}

We have addressed the long-standing problem of finding a qubit
formula for the relative entropy of entanglement \cite{Eisert} or,
equivalently, of finding the CSS $\sigma$ for a given entangled
state $\rho$. We have obtained a solution to the inverse problem
by finding a compact expression for an entangled state $\rho$ for
a given CSS $\sigma$, which is a crucial simplification of the
former solution \cite{Ishizaka03}. The usefulness of our formula
can be demonstrated by finding the REE for some special states but
also by analyzing general properties of the REE. Thus, we have
studied (i) weak and strong additivity of the REE, (ii) how the
REE is related to the Rains upper bound for the entanglement of
distillation, and (iii) nonuniqueness of the closest separable
states for Bell states.

All the examples of entangled states $\rho$ with analytical
expression for the CSSs $\sigma$, discussed in Refs.
\cite{Vedral97a,Vedral98,Verstraete01,Verstraete01jpa,Verstraete02,Audenaert02,MG04,Wei04},
can easily be explained by using our formula as follows. By
starting from some special $\sigma$ one should generate $\rho$ and
then try to find an inverse analytical relation to express
$\sigma$ in terms of $\rho$. Thus, for example, we have derived
the well-known formulas for pure, Bell-diagonal states, the
Horodecki states \cite{Horodecki-book}, the Vedral-Plenio states
\cite{Vedral98}, and the Gisin states \cite{Gisin96} but also
obtained new formulas for some other entangled states.

We have analyzed more general states $\RHO{Z}$ with elements
$\{r_i\}$, which can be generated from the CSS $\SIGMA{Z}$ with
elements $\{R_i\}$ via Eqs. (\ref{N32}). The point is that, apart
from some special cases including $R_2=R_3$ and $R_1=R_4=0$, the
set of Eqs. (\ref{N32}) for $\{r_i\}$ as a function of $\{R_i\}$
seemingly cannot be solved for $\{R_i\}$ due to the presence of
logarithmic functions (\ref{N25}). Thus, we cannot express
$\sigma_Z$ in terms of $\rho_Z$ in general. Although this is not a
proof of impossibility, our analysis of the formula for $\RHO{Z}$
in terms of $\SIGMA{Z}$ strongly suggests that the inverse
problem, which corresponds to finding a compact-form relation for
$\sigma$ in terms of a given $\rho$, can be solved in some special
cases only.

\end{document}